# Packing fraction of clusters formed in free-falling granular streams based on flash X-ray radiography

Yuuya Nagaashi[1*], Akiko M. Nakamura[1], Sunao Hasegawa[2], and Koji Wada[3]

[1] Graduate School of Science, Kobe University, 1-1 Rokkodaicho, Nada-ku, Kobe 657-8501, Japan

[2] Institute of Space and Astronautical Science, Japan Aerospace Exploration Agency, 3-1-1 Yoshinodai, Chuo-ku, Sagamihara, Kanagawa 252-5210, Japan

[3] Planetary Exploration Research Center, Chiba Institute of Technology, 2-17-1 Tsudanuma, Narashino, Chiba 275-0016, Japan

*y.nagaashi@stu.kobe-u.ac.jp

**Abstract**

We study the packing fraction of clusters in free-falling streams of spherical and irregularly shaped particles using flash X-ray radiography. The estimated packing fraction of clusters is low enough to correspond to coordination numbers less than 6. Such coordination numbers in numerical simulations correspond to aggregates that collide and grow without bouncing. Moreover, the streams of irregular particles evolved faster and formed clusters of larger sizes with lower packing fraction. This result on granular streams suggests that particle shape has a significant effect on the agglomeration process of granular materials.

**I. INTRODUCTION**

Understanding the agglomeration of particles is crucial to various settings ranging from protoplanetary discs [1-6], planetary rings [7, 8], cirrus clouds [9] to powder technology [10]. Cluster formation in free-falling granular streams has been examined experimentally using high-speed camera imagery, mainly for spherical particles of different diameters and compositions under various ambient pressure and cohesion [11-13]. They showed that clustering in a granular stream is driven by cohesive force due to van der Waals interaction or capillary force between surface asperities. Numerical simulations, which studied the effect of the cohesion and the restitution coefficient of particles on clustering, showed that the inelasticity affects the degree to which a stream remains collimated and qualitatively reproduced the formation of clusters or droplets in a stream consisting of spherical particles with the cohesive force of hundreds of nano-Newtons [14]. Their formation was intimately associated with spatial heterogeneities in the relative particle velocities, and regions of larger relative velocity deform into necks, and the frozen clusters or droplets are formed as the separation process is complete [14]. The cluster formation is accounted for the energy loss during collisions between cohesive particles within; thus, a granular stream can effectively act as a sensitive probe for minute cohesion in granular systems [12, 14].

Numerical simulations have shown that the larger the particles' adhesion, the higher the average number of particles in a cluster, with a contact number of approximately 3 to 4 [14]; however, experimental observation of clusters using a high-speed camera cannot render such information. In this study, we experimentally measure the packing fraction of the clusters using flash X-ray radiography, which can freeze the particle movement owing to its short exposure time of less than 30 ns. Packing fraction of clusters is one of the important physical properties for determining their interaction with gas [6] and their collisional characteristics [15, 16]. As a consequence, we find that the clusters of irregularly shaped particles have a larger number of particles and a lower packing fraction.

**II. EXPERIMENTAL METHOD**

The experimental configuration is shown in Figure 1(a). The particles had a free fall as a granular stream after opening the electromagnetic shutter at the bottom of a plastic funnel with a circular aperture of 12 mm containing sample particles [13]. The particles consisted of spherical glass beads and irregularly shaped glass powders of density 2500 kg/m$^3$, as shown in Fig. 1(b) and 1(c). Based on optical microscope images (resolution: 1.3 µm/pixel), the axial ratio $b/a$ ($a$ and $b$: major and minor axes, respectively, in the elliptical fitting of the projected area) of beads and powders are $0.97 \pm 0.06$ and $0.58 \pm 0.17$, respectively [17]. The size distribution of both types of particles is shown in Fig. 1(d), and the median size is 45 µm. The particles of this size were chosen because they are easy to handle and form clusters in granular streams [13]. The experiments were carried out in a vacuum chamber (diameter: 1.5 m; height: 2.0 m), and the ambient pressure in the chamber was set to $10^4$ Pa. While a previous study found no effect of ambient pressure (30–$10^5$ Pa) on clustering [12], another previous study conducted in the ambient pressure of 80–$10^5$ Pa found that clustering became difficult to occur under reduced pressure [13]. In this study, we chose $10^4$ Pa.

A 300 KV flash X-ray radiography system (Pulserad model 43733A) was used to capture transient morphology of the granular stream. Unlike the observations in previous studies [12, 18] in which high-speed cameras were dropped, the position of the X-ray source was fixed in this study. Instead, the position of the funnel was set at five different altitudes, ranging between 0 m and 1.35 m, measured from the center of the X-ray beam. The X-ray source radiated in the horizontal direction, perpendicular/normal with respect to the vertical direction, i.e., the direction of the granular stream. An imaging plate was set aside from the stream to detect the flash X-ray intensity at the opposite side of the flash X-ray source across the granular stream. Here, we define the *x-y-z* coordinate as follows: the *z*-axis is the vertical direction along the granular stream; the *y*-axis is along the line of sight of the imaging equipment; and the *x*-axis is perpendicular with respect to the *y* and *z* axes. The origin of the *x-y* coordinate corresponded to the center of each cluster, and the origin of the *z* coordinate was at the opening of the funnel. X-ray intensity data were digitized into an image of 4000 × 8000 pixels with a 16-bit brightness level; however, the diameter of the beam was ~3300 pixel, corresponding to a typical spatial resolution of ~45 µm/pixel, and the maximum intensity was ~3000 counts, i.e., signal-to-noise ratio (S/N) less than ~55.

For an incident X-ray intensity of $I_0$, the X-ray intensity after passing through the granular stream *I* is given by the depth of the granular stream along the line of sight $l(x,z)$, and the packing fraction inside the cluster, $\varphi(x,y,z)$. Assuming a circular cross-section for the clusters in the *x-y* plane because the granular stream flows out from the circular opening of the funnel, the intensity ratio is as follows:

$$\frac{I(x,z)}{I_0(x,z)} = e^{-\kappa \Sigma(x,z)} \quad (1)$$

$$\Sigma(x,z) = \int_{-l(x,z)/2}^{l(x,z)/2} \rho_p \varphi(x,y,z) dy \quad (2)$$

$$l(x,z) = 2\sqrt{w(z)^2 - x^2} \quad (3)$$

Here, $\kappa$ is a mass absorption coefficient that depends on the particle composition, which was determined by a calibration experiment, $\Sigma(x,z)$ is column density, $\rho_p$ is particle mass density, and $2w(z)$ is the stream width. For the calibration experiment, we acquired flash X-ray images of the layers of the spherical glass beads used in this study with known column density. We obtained $\kappa = 0.0241 \pm 0.0003$ m$^2$/kg. We assume a constant packing fraction in the *x-y* plane for simplicity.

**III. RESULTS AND DISCUSSION**

Figure 2(a)-2(f) shows the results of flash X-ray imaging. After falling as glass powder, the streams of spherical particles and irregular particles became constricted, and this was followed by the formation of clusters. The stream of irregular particles showed rapid and easy cluster formation, which had a larger width than that of spherical particles.

To estimate the packing fraction of the streams $\varphi$ using Eq. (1-3), we choose data in the *x*-direction through a cluster and the reference intensity data (3 pixels wide in the z-direction and in the *x*-direction, without the influence of clusters). The spatial pattern of X-ray irradiation was corrected by normalizing the data of the clusters with respect to the reference intensity data. For these X-ray intensity profiles, the packing fraction $\varphi$ and the half-width $w(z)$, minimizing the residuals, were calculated. Examples of the results are shown in Fig. 3(a) and (b). The scatter of the data is consistent with the S/N ratio of the flash X-ray.

The packing fraction was estimated at 1 cm vertical intervals for the part of the granular stream near the aperture of the funnel. However, when clustering occurred, and the granular stream became discrete, the packing fraction was estimated at the altitude near the center of the clusters. The evolutions of the packing fraction of the granular streams are shown in Fig. 4. The scatters in this figure suggest that there is a range of packing fractions of clusters. There was a large difference between the packing fractions for spherical particles upon flow initiation and that after the completion of clustering. However, the difference was small, and the evolution process of packing fraction is likely to be completed more quickly for irregular particles. In spherical particles, the reduction in the packing fraction was nearly completed before the fall distance reached 0.7 m. This was prior to the cluster formation when the stream of spherical particles became discrete in the vertical direction (Fig. 2), and the packing fraction of the streams ceased to evolve further. These results are consistent with the fact that the numerical simulation of a granular stream of spherical particles evolved actively just after flowing out and completed the evolution inside the clusters prior to the completion of cluster formation [14].

In order to estimate the average packing fraction of the clusters, the data are fitted by,

$$\varphi = (\varphi_0 - \varphi_f) \exp(-z/z_\varphi) + \varphi_f, (4)$$

where $\varphi_0$, $\varphi_f$, and $z_\varphi$ are constants. We considered the initial packing fraction $\varphi_0$ as the fraction in the funnel prior to the flow of $0.576 \pm 0.012$ and $0.308 \pm 0.004$ for the spherical particles and the irregular particles, respectively. As a consequence, the average final-packing fraction $\varphi_f$ was $0.34 \pm 0.01$ and $0.21 \pm 0.01$, and the characteristic fall distance $z_\varphi$ was $0.12 \pm 0.03$ and $0.045 \pm 0.015$ m, respectively.

Given the elapsed time since the particles leave the funnel is $t$, the relative velocity between particles in the stream is $v_{rel}(t)$, and the effective restitution coefficient in the inter-particle collisions is constant at $\varepsilon$, the relative velocity after one collision is denoted by $\varepsilon v_{rel}(t)$. Given the particle radius as $r_p$, and the mass density of the stream as $\rho(t)$, the mean free time $\tau(t)$ in the stream is as follows:

$$\tau(t) \sim \frac{4}{3} \frac{\rho_p}{\rho(t)} \frac{r_p}{v_{rel}(t)}. \quad (5)$$

Thus,

$$\frac{dv_{rel}(t)}{dt} \sim \frac{(\varepsilon - 1)v_{rel}(t)}{\tau(t)} \sim -\frac{(1-\varepsilon)3\rho(t)}{4\rho_p r_p} v_{rel}(t)^2 = -\frac{(1-\varepsilon)3\varphi(t)}{4r_p} v_{rel}(t)^2, \quad (6)$$

where $\varphi(t)$ denotes $\rho(t)/\rho_p$.

As shown in Fig. 4, the packing fraction varies only by a factor, so for simplicity, we assume that $\varphi = \varphi_0$, and $v_{rel}(t)$ is given as follows:

$$v_{rel} = \frac{v_{rel0}}{1 + \frac{t}{\tau_0}} = v_{rel0} \left( 1 + \sqrt{\frac{2z}{g\tau_0^2} + \left(\frac{v_0}{g\tau_0}\right)^2} - \frac{v_0}{g\tau_0} \right)^{-1}, \quad (7)$$

$$\tau_0 = \frac{4}{3(1-\varepsilon)} \frac{1}{\varphi_0} \frac{r_p}{v_{rel0}}, \quad (8)$$

where $v_0$ is the downward outflow velocity from the funnel, $g$ is the gravitational acceleration, and $v_{rel0}$ is the particle relative velocity at $t = 0$ (i.e., at $z = 0$), which is ~0.01 m/s [13]. For $\frac{2z}{g\tau_0^2} \ll 1$, Eq. (7) is rewritten as follows:

$$v_{rel} \sim v_{rel0} \left(1 - \frac{z}{v_0 \tau_0}\right) \sim v_{rel0} e^{-\frac{z}{v_0 \tau_0}}. \quad (9)$$

For spherical particles, given that $\varepsilon_s = 0.98$ (assuming surface energy of 0.025 J/m² [20], Young's modulus of 73 GPa and a Poisson's ratio of 0.17 [21] of silica, and the Johnson-Kendall-Roberts model [22]), and $v_0 = 0.25$ m/s (based on an empirical equation [23]), the characteristic distance $z_0 = v_0 \tau_0$ is 0.065 m. Taking the decrease of packing fraction with distance into account, the decay time of relative velocity and the characteristic distance would increase.

If $v_{rel0}$ and $v_0$ do not depend on the shape of the particles, the relationship between the effective restitution coefficients of spherical and irregular particles, $\varepsilon_s$ and $\varepsilon_i$, respectively, can be written using the initial packing fraction $\varphi_{0,s}$ and $\varphi_{0,i}$ and the characteristic distance of packing fraction decay $z_{\varphi,s}$ and $z_{\varphi,i}$ of spherical and irregular particles, respectively, as follows:

$$\frac{1-\varepsilon_i}{1-\varepsilon_s} = \frac{\varphi_{0,s}z_{\varphi,s}}{\varphi_{0,i}z_{\varphi,i}} \sim 5, (10)$$

where we used $\varphi_{0,s} \sim 0.58$, $\varphi_{0,i} \sim 0.31$, $z_{\varphi,s} \sim 0.12$ m, and $z_{\varphi,i} \sim 0.045$ m. Eq. (10) suggests that the effective restitution coefficient of the irregular particle is smaller than those of spherical particles. The lower effective restitution coefficient of irregular particles is consistent with a previous study [2], in which it was suggested that irregularly shaped particles have higher critical collision velocities because they lose more energy in multiple contacts during a single encounter. The quantitative clue of the effective restitution coefficient of irregular particles shown here will be useful in the understanding of the development of structures in ejecta curtain formed by an impact on planetary regolith [24] and dense planetary rings [25].

The evolution of the width of the granular stream derived simultaneously with the packing fraction [Fig. 4] by the fittings [Fig. 3(a) and 3(b)] is shown in Fig. 5. The streams of irregular particles formed the clusters with various widths, and as with the packing fraction evolution, they evolved more rapidly. The cluster width at a fall distance of 1.1–1.4 m is $5.0 \pm 0.5$ mm and $7.4 \pm 1.3$ mm for spherical and irregular particles.

Given the distance $\lambda(t)$ between a particle that begins to fall freely from the funnel ($z = 0$) at a downward velocity $v_0$ at time $t = 0$, and a particle that begins to fall prior and is free-falling at time $t = 0$ at $z = \lambda_0$ with downward velocity $\sqrt{2g\lambda_0 + v_0^2}$, $\lambda(t)$ can be written as follows [11]:

$$\lambda(t) = \left(\lambda_0 + \sqrt{2g\lambda_0 + v_0^2}\, t + \frac{1}{2}gt^2\right) - \left(v_0 t + \frac{1}{2}gt^2\right). (11)$$

Assuming mass conservation, the half-width $w(t)$ and packing fraction $\varphi(t)$ of the stream have the following relationship:

$$w(t)^2 \lambda(t) \rho_p \varphi(t) = const. (12)$$

Thus,

$$w = w_0 \left(\frac{\lambda}{\lambda_0}\right)^{-\frac{1}{2}} \left(\frac{\varphi}{\varphi_0}\right)^{-\frac{1}{2}}. (13)$$

Curves obtained by substituting Eqs. (4) and (11) into Eq. (13) are shown in Figure 5. Here we use $\lambda_0 = 2r_p$. The experimental results are approximated by Eq. (13) until the clusters are formed.

With the estimated packing fraction and the width of the clusters, the network sizes of the clusters of spherical particles and irregular particles could be roughly compared. We assumed ellipsoids with a 2:1:1 axial ratio based on the previously reported ratio of the cluster's length along the stream and the width perpendicular to the stream [12, 13]. Further, using the width and the packing fraction of 13 clusters of spherical particles at a fall distance of 1.1–1.4 m [Fig. 4 and 5], the average number of spherical glass particles constituting the clusters was estimated to be $(9.3 \pm 2.8) \times 10^5$. Meanwhile, for 9 clusters of irregular particles at a similar fall distance, the average number of constituent irregular particles, for which monomer mass was assumed to be the same as the spherical particle, was found to be $(1.9 \pm 1.0) \times 10^6$, which is about twice as large as that of spherical particles. However, the assumption of the ellipsoidal clusters with a constant packing fraction can result in overestimation of the absolute number of particles. Integrating the column densities over both profiles shown in Fig. 6(a) and 6(b), the total numbers of constituent particles of the two clusters of spherical particles is $9 \times 10^5$, and the cluster of irregular particles is $3 \times 10^6$, respectively. It gives a value about half of that of approximated from the ellipsoidal clusters: those of the ellipsoidal approximations are $2.3 \times 10^6$ (sum of two clusters) and $4.7 \times 10^6$, respectively. The numerical simulation using spherical particles has shown that the network size of the clusters is directly associated with the cohesive strength of the particles [14]. Drawing an analogy, the fact that irregularly shaped particles agglomerate easily could affect not only in the rate of cluster formation but also the network size.

According to empirical relationships between the coordination number and the packing fraction for aggregates of spherical particles [15, 16, 26] (Fig. 7), the packing fraction of ~0.3 corresponds to an average coordination number of ~3–5, which is consistent with numerical results [14]. Meanwhile, the average coordination number for irregular particles is estimated to be ~2–4 on the basis of the empirical relationship for spherical particles, which is a new information regarding the cluster formation in a granular stream. Coordination number is an important parameter in the context of dust growth in protoplanetary disks. Numerical simulations of dust aggregates consisting of spherical particles have shown that dust aggregates would bounce during the collisions only if their coordination numbers are greater than 6 [15]. Knowledge about whether bouncing is present or absent is important to assess whether the dust in the protoplanetary disks can or cannot grow due to the bouncing barrier [27, 28]. The particles in the clusters formed in the granular stream are considered to have coordination numbers less than the bouncing threshold of 6, which implies that the clusters in the granular streams can naturally simulate dust growth in protoplanetary disks. In addition, the packing fraction also affects the tensile strength of dust aggregates [29], which should affect the fragmentation conditions. Thus, the packing fraction of the clusters provide important insights into discussions of the dust growth related to planet formation [30] and the size distribution of dense planetary rings [8].

The evolution of a granular stream not only depends on the cohesion between the particles, but also on the particle shape. Given that naturally occurring particles are generally non-spherical, there is a need to actively incorporate irregularly shaped particles in the study of agglomeration phenomena of particles, including numerical simulations. Furthermore, the tendency of particle aggregation is manifested and thus can be evaluated from the cluster formation rate and the network size.

**Acknowledgements**

This research was supported by the Hypervelocity Impact Facility (former facility name: the Space Plasma Laboratory), ISAS, JAXA and JSPS KAKENHI (No. 19H05081). Y. N. is grateful to the Hosokawa Powder Technology Foundation for their financial support.

**Figures**

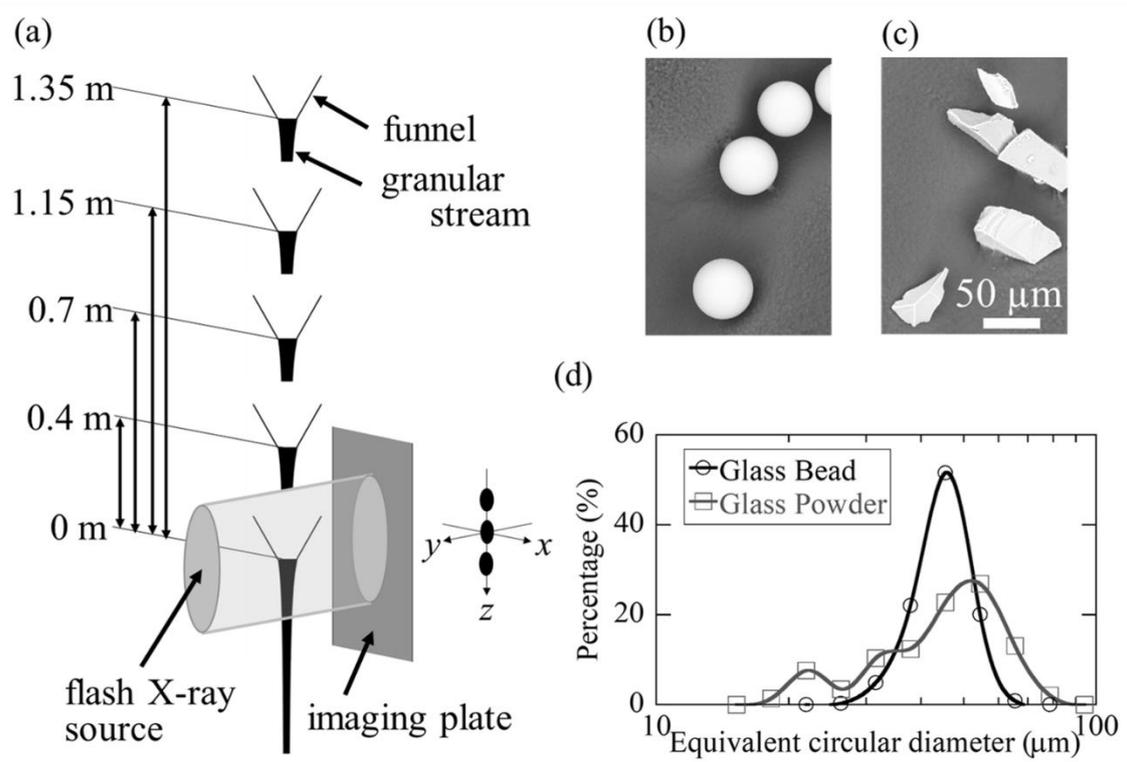

FIG. 1. Flash X-ray imaging of the granular streams. (a) Schematic diagram of the experimental configuration and definition of *x-y-z* coordinate in this study. (b-c) Scanning electron microscopy (SEM) images of (b) the glass beads and (c) the glass powders [17]. (d) Particle size distributions [17].

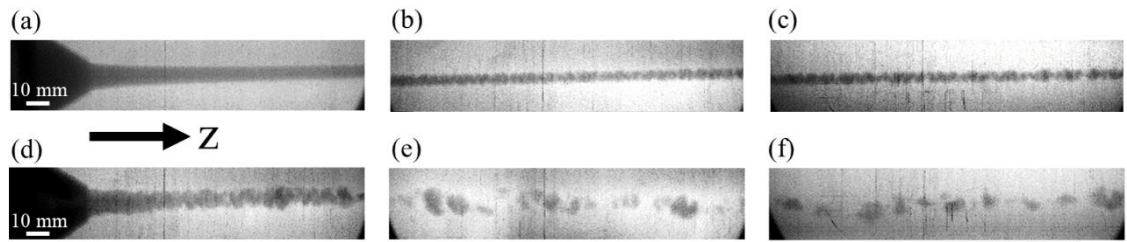

FIG. 2. Flash X-ray transmission images of the streams [19]. Streams of spherical glass particles at fall distance (a) z = 0, (b) 0.4, and (c) 0.7 m, respectively, and streams of irregularly shaped glass particles at fall distance (d) z = 0, (e) 0.4, and (h) 0.7 m, respectively. Left is up. The scars on the image are due to mechanical scratches on the imaging plate.

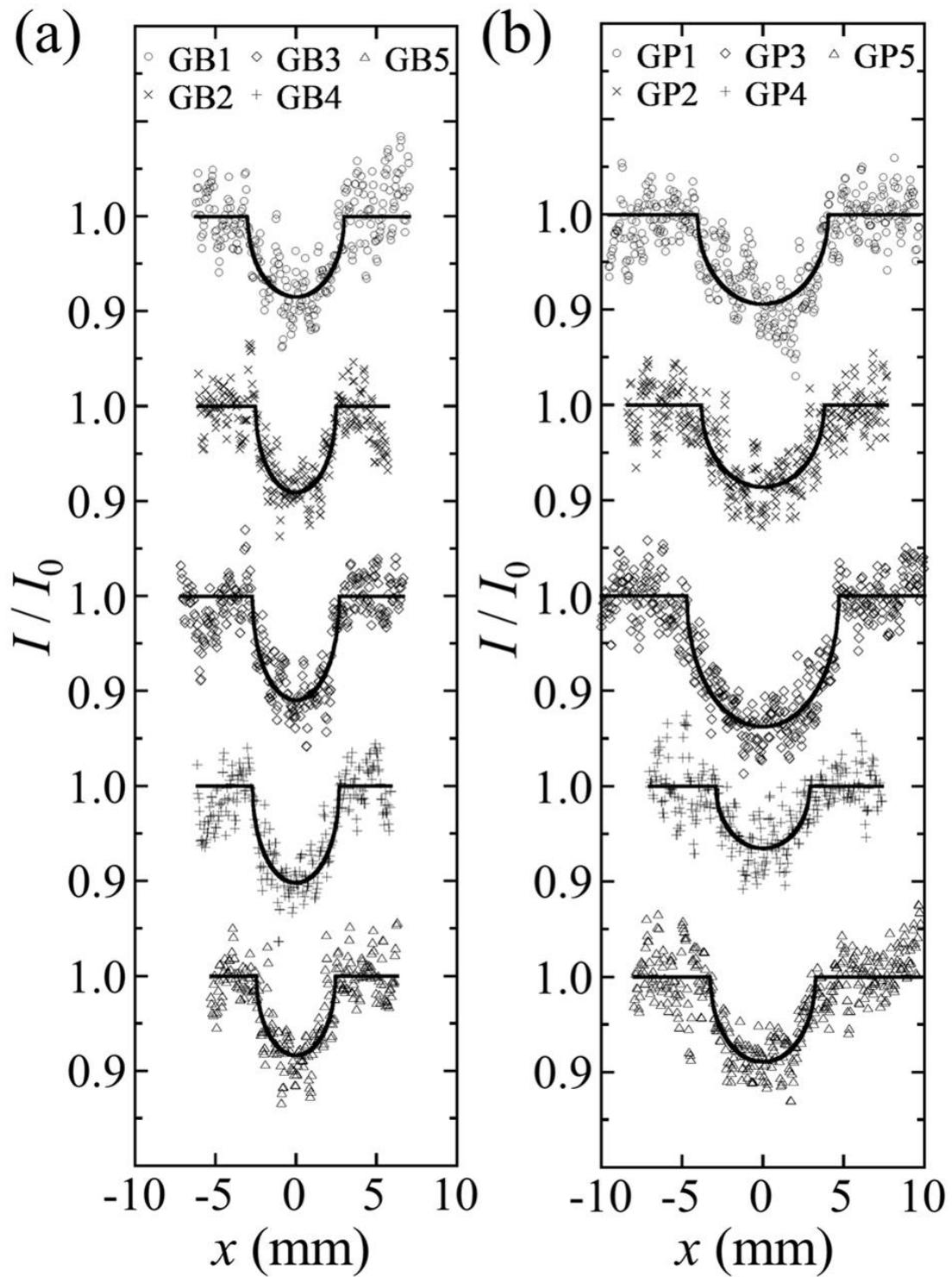

FIG. 3. Comparisons of the X-ray profiles of clusters and the fitting curves based on assumptions of constant packing fraction and axial symmetry. (a) Clusters of spherical particles (GB1-GB5 in the order from the top) and (b) irregular particles (GP1-GP5 in the order from the top), observed for a fall distance of 1.1–1.2 m.

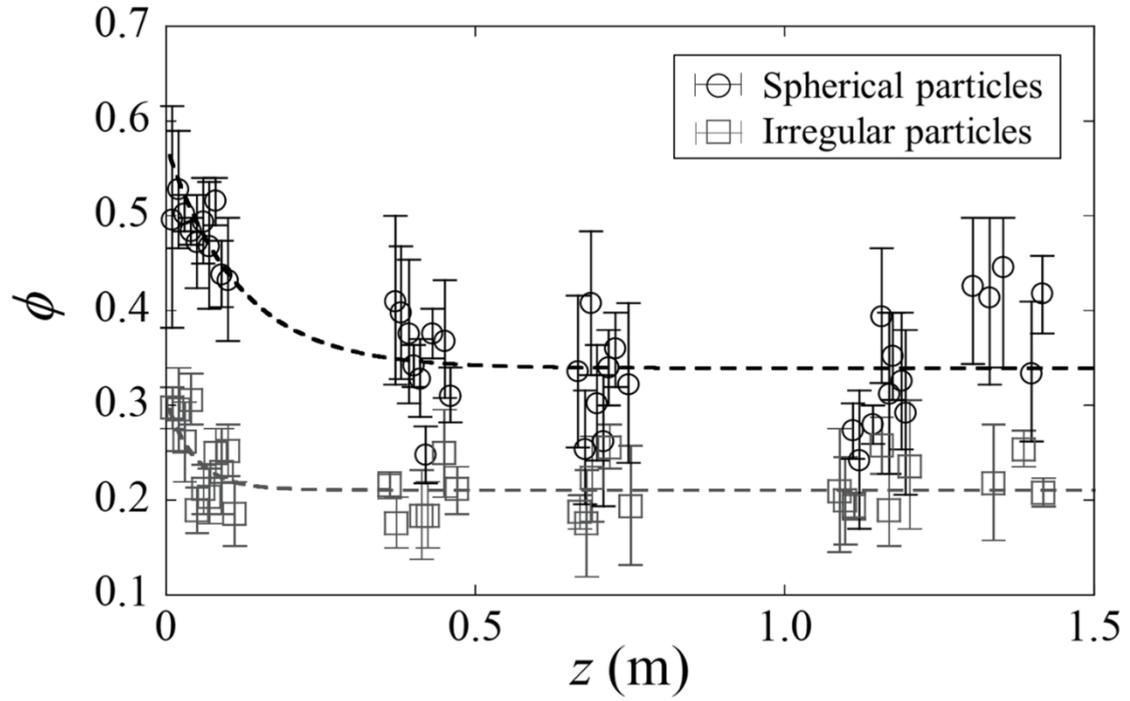

FIG. 4. The vertical evolution of the packing fraction $\varphi$ of the granular stream. The error bars are the uncertainties corresponding to the root mean squares of the residuals in FIG. 3(a) and 3(b). Each dashed curve is obtained by plotting the expression in Eq. (4).

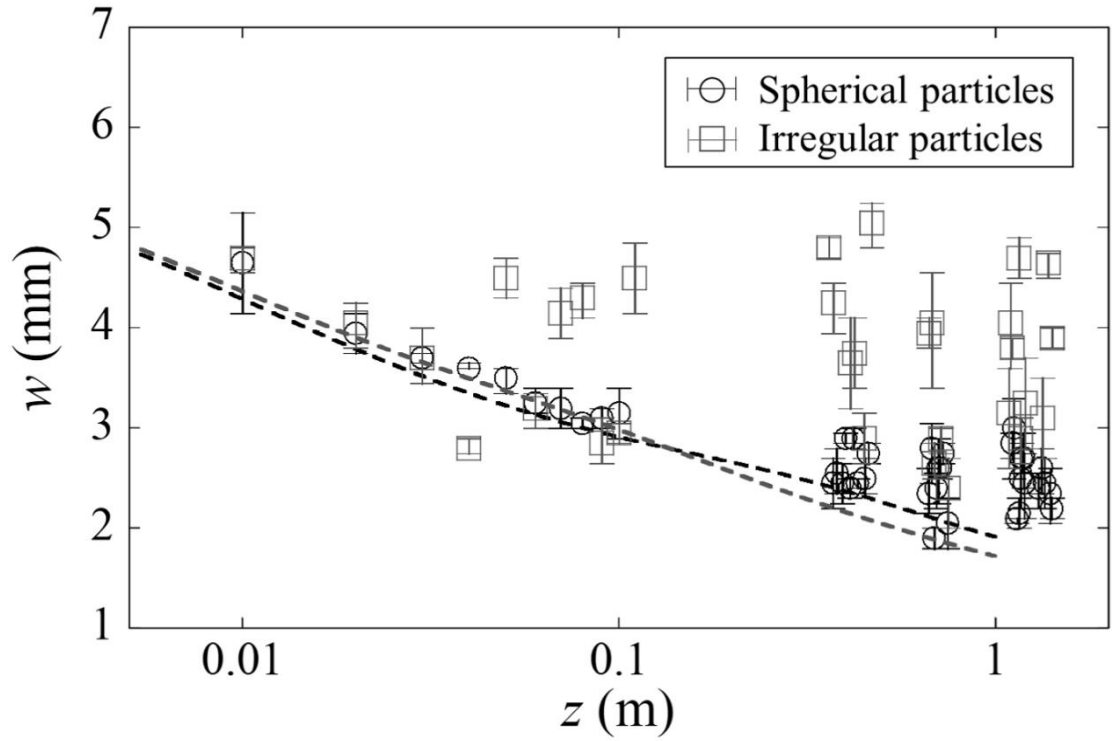

FIG. 5. The vertical evolution of the half-width $w$ of the granular stream. The error bars are the uncertainties corresponding to the root mean squares of the residuals in FIG. 3(a) and 3(b). Each dashed curve is obtained by plotting the expression in Eq. (13).

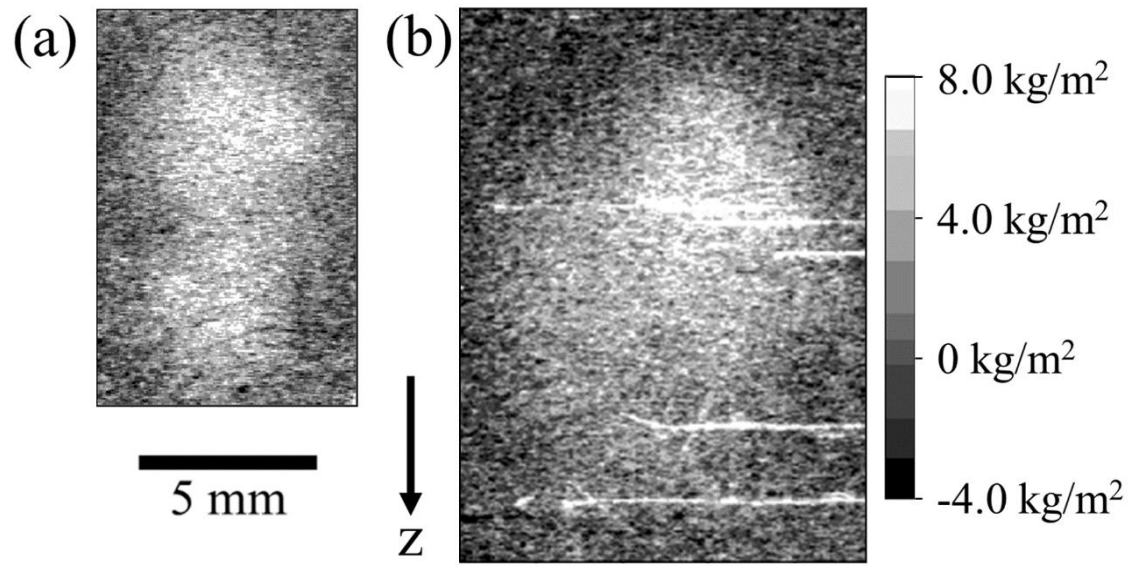

FIG. 6. Column density Σ(x, z) profile of the clusters. The clusters of spherical particles at a fall distance (a) 1.18 m and (b) irregular particles at a fall distance 1.16 m.

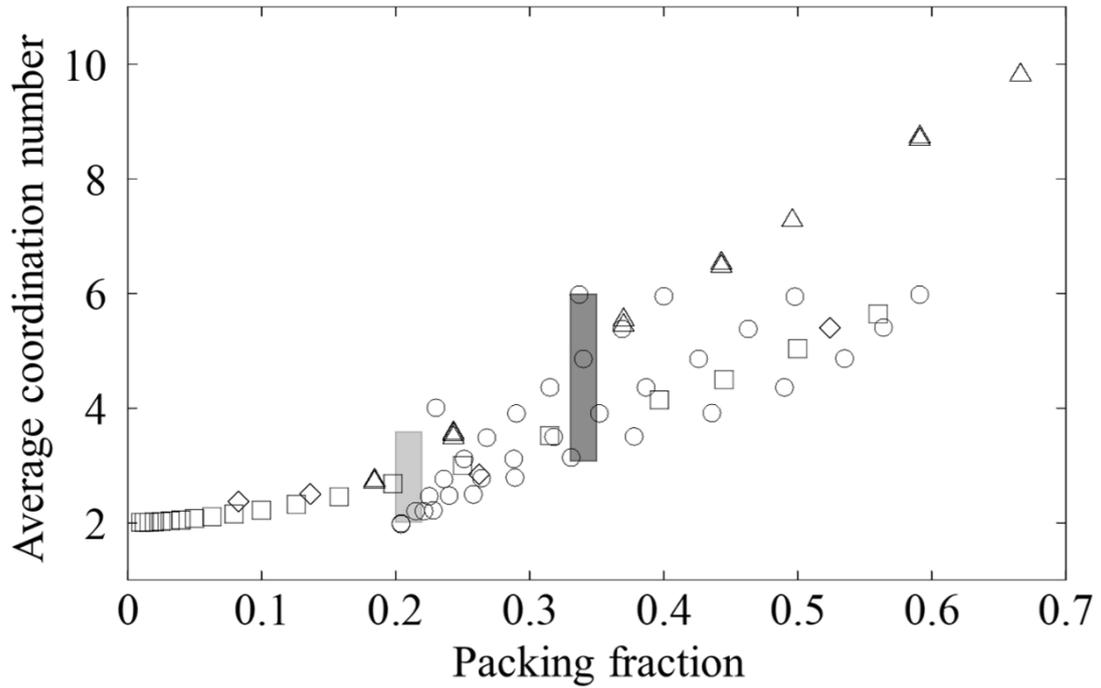

FIG. 7. Coordination numbers of the clusters. The relationship between the packing fraction and the coordination number of the dust aggregates composed of spherical particles derived in the numerical simulations is shown by circles (ballistic agglomeration with migration [15,16]), diamonds (cubic lattice [15]), triangles (close-packed and particle-extracted [15]), and squares (ballistic cluster-cluster aggregation [26]). The range of packing fraction of the clusters of the spherical particles and irregular particles is indicated by deep gray and light gray patches, respectively.